# Three-dimensional Mid-air Acoustic Manipulation by Ultrasonic Phased Arrays

**Authors and Affiliations**
Yoichi Ochiai[*1], Takayuki Hoshi[*2], and Jun Rekimoto[*1,3]
[1] The University of Tokyo, Graduate School of Interdisciplinary Information Studies
 7-3-1 Hongo, Bunkyo-ku, Tokyo, 113-0033 Japan
[2] Nagoya Institute of Technology
 Gokisocho, Showa-ku, Nagoyashi, Aichi-ken, 466-8555 Japan
[3] Sony CSL
 3-14-13 Higashigotanda, Shinagawa-ku Tokyo 141-0022 Japan

**Abstract**
The essence of levitation technology is the countervailing of gravity. It is known that an ultrasound standing wave is capable of suspending small particles at its sound pressure nodes. The acoustic axis of the ultrasound beam in conventional studies was parallel to the gravitational force, and the levitated objects were manipulated along the fixed axis (i.e. one-dimensionally) by controlling the phases or frequencies of bolted Langevin-type transducers. In the present study, we considered extended acoustic manipulation whereby millimetre-sized particles were levitated and moved three-dimensionally by localised ultrasonic standing waves, which were generated by ultrasonic phased arrays. Our manipulation system has two original features. One is the direction of the ultrasound beam, which is arbitrary because the force acting toward its centre is also utilised. The other is the manipulation principle by which a localised standing wave is generated at an arbitrary position and moved three-dimensionally by opposed and ultrasonic phased arrays. We experimentally confirmed that expanded-polystyrene particles of 0.6 mm and 2 mm in diameter could be manipulated by our proposed method.

**Introduction**
Ultrasonic levitation method has been used to levitate lightweight particles[1], small creatures[2], and water droplets[3]. The principle of acoustic levitation was mathematically explained by Gor'kov[4] and Nyborg[5]. The potential energy $U$ of an ultrasound standing wave is given by

$$U = g(x,y)^2 \frac{A^2}{\rho_0 c^2} \left\{ -B + (B+1-\gamma)\cos^2(2\pi \frac{z}{\lambda}) \right\} \quad (1)$$

The acoustic axis coincides with the $z$ axis, and $g(x,y)$ is the normalised cross-sectional distribution of the velocity potential. $A$ is the RMS amplitude; $B$ is given by $3(\rho - \rho_0)/(2\rho + \rho_0)$, where $\rho$ and $\rho_0$ are the densities of a small sphere and the medium, respectively; $\gamma$ is given by $\beta/\beta_0$, where $\beta$ and $\beta_0$ are the compression ratios of the small sphere and the medium, respectively; $c$ is the speed of sound in the medium; and $\lambda$ is the wavelength of ultrasound. The force **F** acting on a sphere of volume $V$ is obtained by $\mathbf{F} = -V\nabla U$. This principle has been examined using bolted Langevin-type transducers with fixed acoustic axes.

**Material and Methods**
*Phased Array*
We innovatively employed ultrasonic phased arrays[6] as transducers. Two arrays opposed to each other were used to generate a standing wave at their common focal point. It was theoretically determined that the distribution of the focal point $g(x,y)$ generated by a rectangular transducer array was approximated by the sinc function. The two-dimensional sinc function $\text{sinc}(x,y)$ is defined as follows:

$$g(x,y) = \text{sinc}(2\pi \frac{x}{w}, 2\pi \frac{y}{w}) \quad (2)$$

Here, the two-dimensional sinc function $\text{sinc}(x,y)$ is defined as $\sin(x)\sin(y)/xy$. $w$ is the diameter of the focal point given by $2\lambda R/D$, where $R$ and $D$ are the focal length and side length of the rectangular array, respectively. Figure 1 shows the potential energy distribution based on Eqs. (1) and (2) when $y = 0$. It is assumed here that the sphere is made of polystyrene and the medium is air. Hence, $\rho = 1.0 \times 10^3$ kg/m$^3$, $\rho_0 = 1.2$ kg/m$^3$, $\beta = 2.5 \times 10^{-10}$ Pa$^{-1}$, and $\beta_0 = 7.1 \times 10^{-6}$ Pa$^{-1}$. The figure shows that small spheres gravitate toward the acoustic axis of the ultrasound beam at its nodes.

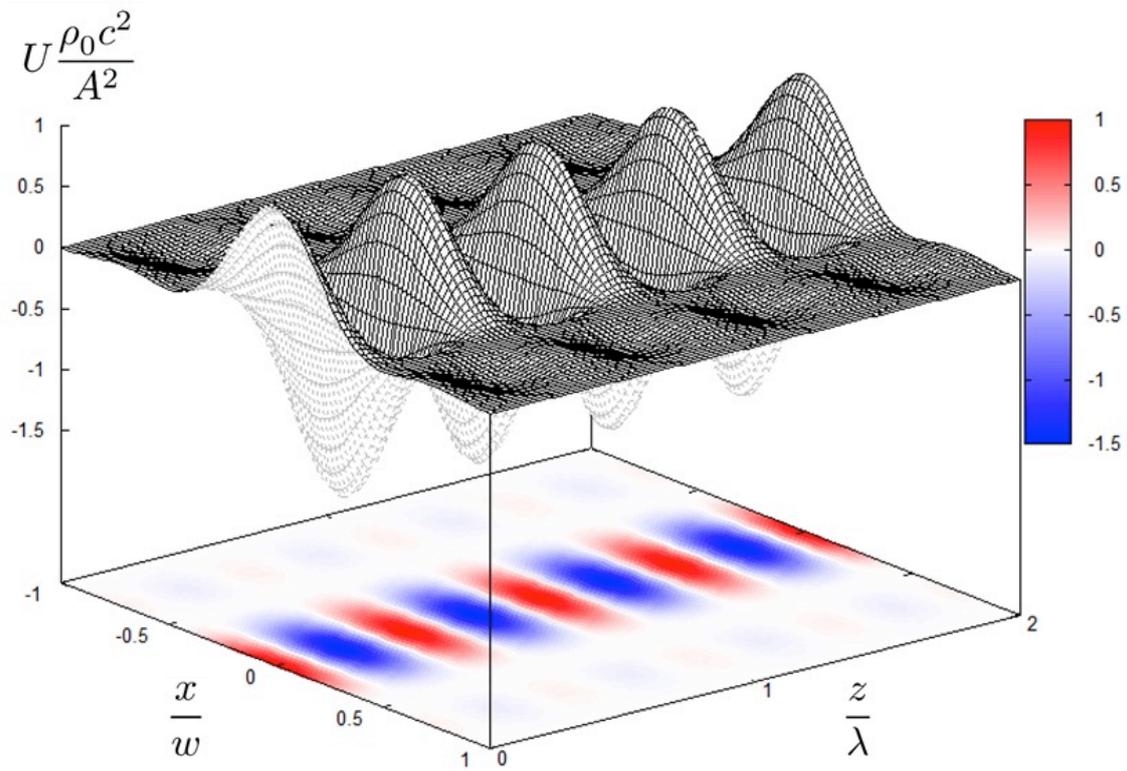

[Figure 1. Potential energy distribution of ultrasonic standing wave.]

The detailed specifications of the phased array (shown in Figure 2) are as follows. It consisted of 285 transducers arranged in a 170 × 170 mm² square area and designed to generate a single focal point by adequate control of their phase differences. The resonant frequency was 40 kHz, and the sound pressure at the peak of the focal point was as high as 2600 Pa (RMS) when the focal length $R$ was 200 mm. The spatial resolution of the position of the focal point was 0.5 mm, and the refresh rate was 1 kHz.

*Levitation Principle*
Here, we discuss how a small polystyrene sphere can be levitated in air by the force per unit volume acting toward the acoustic axis at the nodes. We suppose that $y = 0$ and $z = 0.25\lambda$ (i.e. a node). The $x$ component $F_x/V$ is therefore obtained by determining the gradient of Eq. (1):

$$\frac{Fx}{V} = \frac{4\pi A^2 B}{w\rho_0 c^2} \left\{ \frac{\sin(2\pi \frac{x}{w})\cos(2\pi \frac{x}{w})}{(2\pi \frac{x}{w})^2} - \frac{\sin^2(2\pi \frac{x}{w})}{(2\pi \frac{x}{w})^3} \right\} .(3)$$

A polystyrene sphere can be levitated by this force if its density is less than $F_x/Vg$, where $g = 9.8$ m/s² is the gravitational acceleration. The speed of sound is 340 m/s. The amplitude $A$ generated by two phased arrays has a maximum value of 5200 Pa (RMS) when the focal length $R = 200$ mm, and $w = 20$ mm. $F_x/Vg$ has a peak value of $5.0 \times 10^3$ kg/m³ at $x \approx -0.2w$. This value is greater than the density of polystyrene, and it is therefore expected that a small sphere of polystyrene would be levitated even when the ultrasound beam is perpendicular to the gravitational force.

*Experimental Setup on Stability*
We examined the stability of the manipulation by measuring the duration of the cyclic movement at different frequencies. The test was conducted using two types of particles, namely expanded-polystyrene spheres of diameters 0.6 mm and 2.0 mm. In each trial, a single particle was set at the third node along one of the acoustic axes ($x$ axis) from the intersection of the ultrasound beams. All the directions of movement (i.e. $x$, $y$ along the other acoustic axes, and $z$ perpendicular to the other two axes) were tested. The focal length was set at 260 mm (Figure 2). The sound pressure was set to 70 % of the maximum. The amplitude of the cyclic movement was 15 mm.

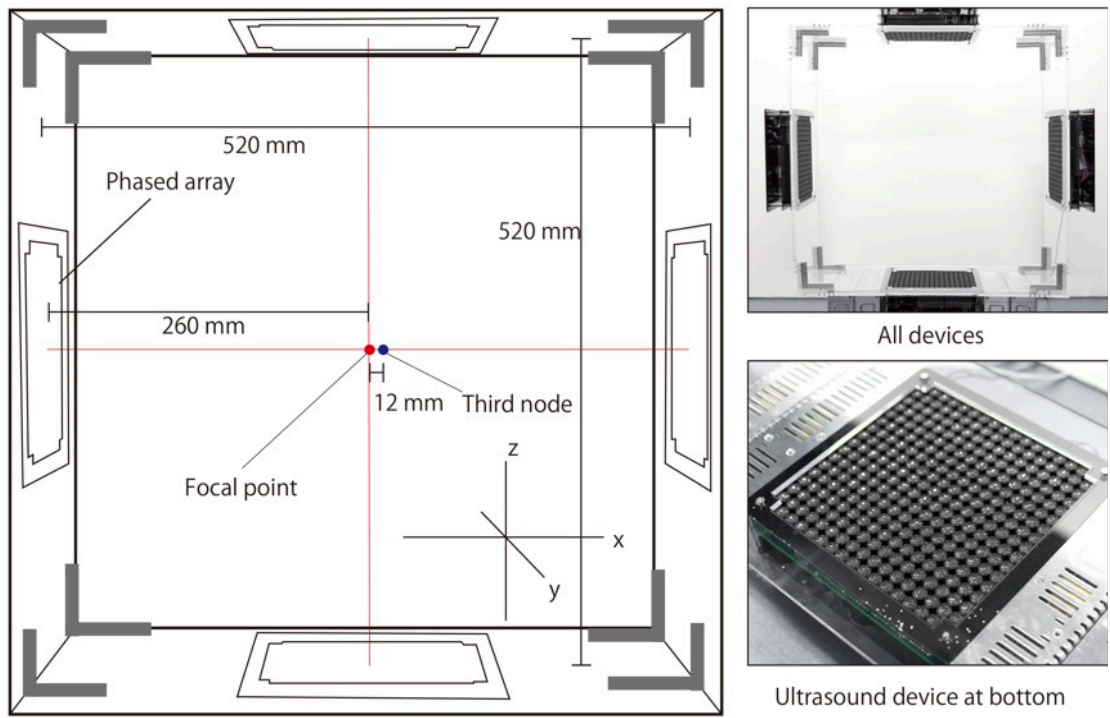

[Figure 2. Illustration and photograph of system setup.]

## Results
### *Levitation and Manipulation*
Multiple ultrasound beams can be overlapped as shown in Figure 3 (left). The expanded-polystyrene particles are trapped at the nodes of both ultrasound beams. The interval between the trapped particles is about 4 mm, which is about half the wavelength of a 40-kHz ultrasound. It can be observed that the particles are more stably levitated using this configuration compared to using a single beam. When the beam moves through a mass of particles, the particles are scooped up and held in the beam as shown in Figure 3 (right).

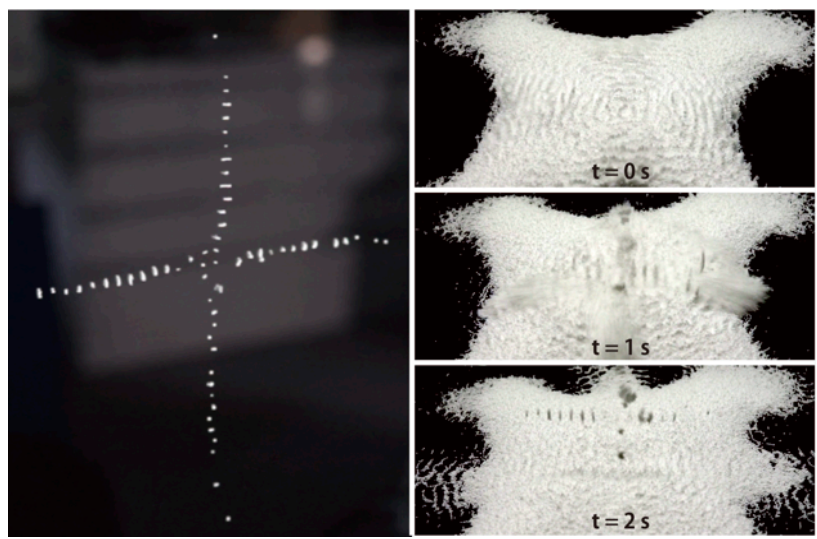

[Figure 3. (left) Levitation and manipulation of particles. (right) Scooping up and holding particles.]

*Stability*
Figure 4 shows the results. The points on the graph indicate the average floating time [s] for the different frequencies [Hz] and the bars indicate the maximum and minimum values. It can be observed that manipulation along the *y* axis was more stable than those along the other axes. We speculate that the manipulations along the *x* and *z* axes tend to induce the discontinuity of ultrasound to change the focal length. Moreover, the graph shows that the 0.6-mm-diameter particles are more stable than the 2.0-mm-diameter particles at higher frequencies. This suggests that larger particles tend to fall from the nodes of a standing wave.

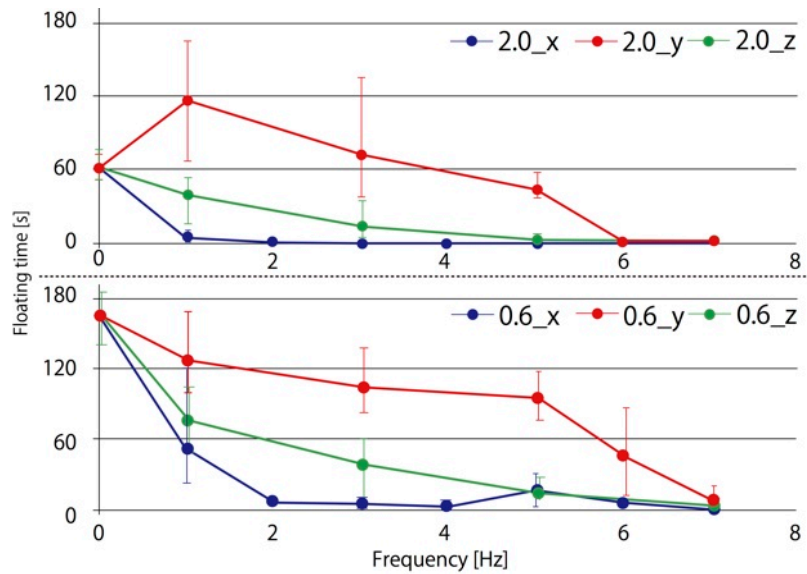

[Figure 4. Results of stability experiments. (top) 2.0 mm particles (bottom) 0.6 mm particles.]

Next, the work space was studied. In the case of movement along one of the acoustic axes, the manipulated particles could approach the ultrasound array to within 60 mm, but fell when they approached nearer. In the case of the movement perpendicular to the acoustic axes, the particles at the more distant nodes fell earlier when they moved away from the centre of the system. A particle at the intersection of the ultrasound beams fell when it came to within 330 mm of the centre.

**Discussion**
There are some factors to be considered in choosing the manipulation target, namely the size and material. The size of the manipulation target is determined by the distribution of the potential energy, and a light material is required. The internal force is also an important factor in selecting the material; for example the electrostatic force determines the maximum number of particles that can be at a single node, and the surface tension of the fluid determines the size of droplets that can be levitated.

**Conclusion**
In conclusion, we have demonstrated an extended acoustic manipulation by which millimetre-sized particles can be levitated and moved three-dimensionally by localised ultrasonic standing waves generated by ultrasonic phased arrays. In addition to the presented examples, we also tested other small objects such as a feather and droplets of alcohol and a colloidal solution.
In a future work, we will use 25 kHz transducers instead of the 40 kHz type. Then the 4-mm node intervals by 40 kHz transducers will be extended to 8 mm. This would enable the manipulation of larger particles.
It has not escaped our notice that our developed method for levitation under gravity suggests the possibility of developing a technology for handling objects under microgravity.